\documentclass[aps,prl,reprint]{revtex4-1}

\usepackage{graphicx}
\usepackage{amssymb}
\usepackage{color}
\usepackage[utf8]{inputenc}
\usepackage{siunitx}
\usepackage{endnotes}
\usepackage{natbib}

\DeclareTextSymbol{\degre}{OT1}{23}
\begin{document}

\title{Vitreous silica distends in helium gas: acoustic vs. static compressibilities}

\author{Coralie Weigel$^{1,2}$, Alain Polian$^{3}$, Mathieu Kint$^{1,2}$, Benoit Ruffl\'e$^{1,2}$, Marie Foret$^{1,2}$,Ren\'e Vacher$^{1,2}$}

\affiliation{$^1$Universit\'e Montpellier 2, Laboratoire Charles Coulomb UMR 5221, F-34095, Montpellier, France\\
$^2$CNRS, Laboratoire Charles Coulomb UMR 5221, F-34095, Montpellier, France\\
$^3$IMPMC-CNRS UMR 7590, Universit\'e P.et M. Curie - Paris 6, B115, 4 pl. Jussieu, F-75252 Paris CEDEX 05, France}

\date\today

\begin{abstract}
Sound velocities of vitreous silica are measured under He compression in the pressure range 0-6~GPa by Brillouin light scattering.
It is found that the well-known anomalous maximum in the pressure dependence of the compressibility is suppressed by He incorporation into the silica network.
This shows that the elastic anomaly relates to the collapse of the largest interstitial voids in the structure.
The huge difference between the static and the acoustic compressibilities indicates that the amount of incorporated helium still increases at 6~GPa.
\end{abstract}

\maketitle

Silica is important for countless reasons. 
In its vitreous form, $v$-SiO$_2$, it finds a considerable number of technical applications. 
Further, it is a prototypal glass-forming material of fundamental interest. 
Thus, $v$-SiO$_2$ has been extensively studied over large pressure ($P$) and temperature ($T$) ranges~\cite{hemley_raman_1986,inamura_transformations_2004,sato_sixfold_coordinated_2008,benmore_structural_2010,ruffle_scaling_2010,sen_observation_2004,trachenko_network_2004}. 
Several anomalous physical properties were discovered. 
In particular, the compressibility $\chi$ does not show the 
$P$ and $T$ dependencies observed in most solids. 
As $P$ is increased at room $T$, $\chi$ first increases to reach a maximum around 2~GPa and it decreases thereafter~\cite{bridgman_effects_1953,schroeder_brillouin_1990}. 
The variation of $\chi$ with $T$ is also anomalous. 
It decreases as $T$ increases above ambient up to the glass transition~\cite{krause_vibrational_1968,polian_elastic_2002,vacher_anharmonic_2005}. 
The relation of these anomalies to the structure of $v$-SiO$_2$ remains debated.

Two different models, supported by numerical simulations, have been proposed. 
In a first one, the changes of $\chi$ with $T$ and $P$ are related to structural modifications associated with rotations of Si$-$O$-$Si bonds in six-membered rings, similar to those occurring in the phase transition of cristobalite~\cite{huang_amorphous-amorphous_2004,liang_mechanical_2007}. 
This was termed a {\em local progressive transition}. 
Alternatively, in a model based on {\em network flexibility}, $\chi$ is maximum in an intermediate range of $P$ over which volume changes can be accommodated by low-energy buckling of the ``floppy'' tetrahedral network~\cite{walker_origin_2007}. 
Both approaches rely on the existence of a sufficiently large free volume allowing strong structural modifications with $T$ and $P$. 
The mass density of $v$-SiO$_2$ is indeed rather small compared to crystal quartz. 
Also, the fact that $v$-SiO$_2$ can be permanently densified by more than 20\% by pressurization at 20~GPa~\cite{susman_intermediate-range_1991,Pol90} indicates that the free volume is unusually large.

The distribution of interstitial voids in glasses can be investigated by studying gas solubility. 
Results for $v$-SiO$_2$ were analyzed assuming a log-normal distribution of voids, with a diameter around 0.2~nm and a width of about 0.1~nm~\cite{shackelford_interstitial_1978,shackelford_gas_1999}, in agreement with numerical simulations~\cite{chan_theoretical_1991}. 
Thermal motions in the glass open the voids to gas atoms. 
A solubility $S\sim$0.0084 mol/cm$^{3}$/GPa was found for He in $v$-SiO$_2$ at $P$ up to 0.13~GPa~\cite{shelby_pressure_1976}.

Recent publications~\cite{shen_effect_2011,sato_helium_2011} revealed that $v$-SiO$_2$ submitted to high $P$ in He atmosphere exhibits a surprisingly small change in volume. 
This was interpreted as a considerable reduction of compressibility~\cite{sato_helium_2011}, leading to an apparent bulk modulus $B=\chi~^{-1}=-V{dP}/{dV} \approx$~110~GPa near ambient $P$. 
This $\chi$ can be called the {\em static} compressibility. 
Another $B$ can be calculated from the longitudinal ($v_{\rm L}$) and transverse ($v_{\rm T}$) sound velocities, and the density $\rho$, all known at ambient $P$, $B~=~\rho(v_{\rm L}^2-\frac{4}{3} v_{\rm T}^2)$. 
One finds an {\em acoustic} modulus $B \approx$~36.5 GPa. 
This large difference in compressibilities must relate to the open structure of $v$-SiO$_2$ allowing He to distend the network in the static limit.

In this Letter, we present Brillouin light-scattering (BLS) measurements of the sound velocity for the longitudinal and transverse acoustic modes as a function of $P$ up to 6~GPa in He atmosphere~\cite{note_pression}.
We show that the elastic moduli are relatively weakly affected by the presence of He, while the minimum in the acoustic bulk modulus is almost completely suppressed.

The samples are Suprasil F300 containing less than 1~ppm OH supplied by Heraeus Quartzglass, Germany. 
We use platelets of 56~$\mu$m thickness and about 100~$\mu$m lateral dimensions. 
The hydrostatic pressure $P$ is applied in a Chervin-type diamond-anvil cell~\cite{chervin_diamond_1995} with diamond culets of 800~$\mu$m diameter. 
The pressure transmitting media are either He or a non-penetrating pressurizing medium (NPPM), namely a methanol-ethanol 4:1 mixture (ME 4:1). 
$P$ is measured by the ruby-fluorescence technique~\cite{chervin_ruby-spheres_2001}.

\begin{figure}
\includegraphics[width=8.5cm]{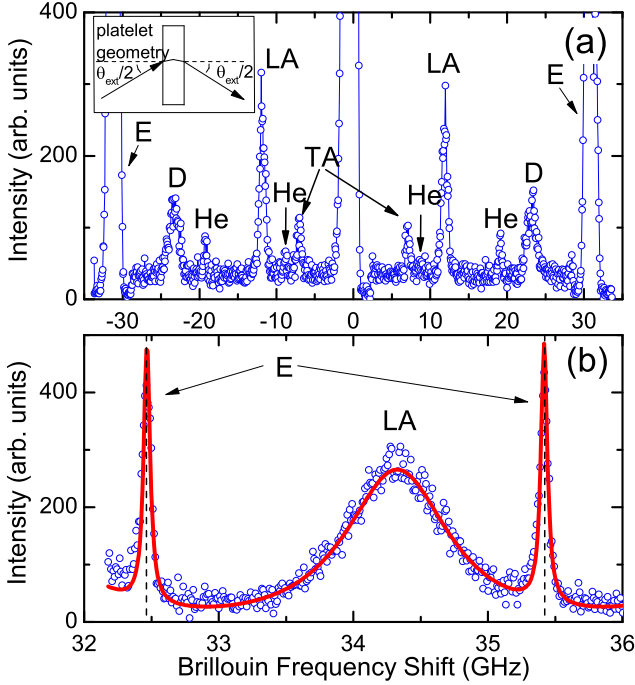}
\caption{(Color online) (a) Brillouin spectrum at 3.4 GPa for $v$-SiO$_2$ in He, obtained with the Sandercock spectrometer. 
The platelet geometry shown in the inset is used. 
As usual, lines emitted by He (noted He in the figure) and by the diamond anvils (D) are observed in addition to the elastic lines (E) and to the transverse and longitudinal Brillouin lines of $v$-SiO$_2$ (TA and LA respectively). 
(b) Brillouin spectrum of the LA modes of $v$-SiO$_2$ at 3.4 GPa measured with the high-resolution spectrometer in the backscattering geometry. 
The two dashed lines locate the periodically transmitted elastic peaks. 
The red line is a fit to the data.}
\label{FigSpectra}
\end{figure}

BLS experiments are performed using the 514.5~nm line of a single frequency argon-ion laser. 
We use two different arrangements. 
The first employs a Sandercock 3+3 pass system using two plane Fabry-Perot's (FP) in tandem~\cite{lindsay_construction_1981}. 
The sample is placed in the platelet geometry with an external scattering angle $\theta_{\rm ext} \approx 60$\degre (see inset in Fig.~\ref{FigSpectra}). 
In this case, the Brillouin frequency shift $\Delta\omega/2\pi$ is given by $\Delta\omega/\omega_0 = (2v/c)\sin \frac{1}{2}\theta_{\rm ext}$, where $\omega_0$ is the angular frequency of the incident light, $v$ the sound velocity, and $c$ the light velocity in vacuum.
This configuration allows transverse and longitudinal modes to be observed on the same spectrum, as shown in Fig.~\ref{FigSpectra}a.

We also use a high resolution spectrometer (HRS) which consists of a plane and a confocal FP in tandem~\cite{sussner_high-precision_1979,rat_anharmonic_2005}. 
This HRS, employed in the backscattering configuration, allows measuring the frequency shifts of longitudinal modes with an accuracy of about 3 MHz. 
In this case, $\Delta\omega$ is given by $\Delta\omega/\omega_0 = (2nv/c)$ where $n$ is the refractive index at the laser wavelength. 
The typical spectrum in Fig.~\ref{FigSpectra}b shows the longitudinal Brillouin line at 3.4 GPa, strongly broadened, and two elastic peaks. 
Owing to the small free spectral range of the confocal FP, the Brillouin peak and the elastic peaks are observed at different interference orders. 
For both series of experiments, the frequency shifts and the linewidths are obtained by fitting the Brillouin lines to a damped harmonic oscillator convoluted with the instrumental profile. 
The profiles are corrected for the broadening introduced by the small sample size~\cite{vacher_finite_2006}, as well as for the finite collection aperture.

\begin{figure}
\includegraphics[width=8.5cm]{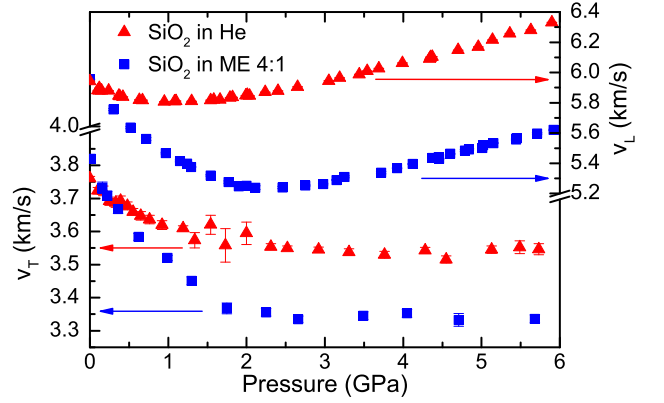}
\caption{(Color online) Transverse and longitudinal sound velocities for $v$-SiO$_2$ as a function of hydrostatic pressures in Helium (red triangles) and in ME 4:1 (blue squares). 
The accuracy of $P$ measurements is 0.1~GPa.}
\label{FigVelocities}
\end{figure}

The Brillouin frequency shift of the longitudinal mode was measured with both spectrometers at room $T$ with $P$ up to 6 GPa. 
Comparing the data at the same $P$ in both geometries gives $n$ of $v$-SiO$_2$ in He, \mbox{$n_{v-\rm SiO2}^{\rm He}(P) =  1.4616 + 0.0081 \times P$}, with $P$ in GPa. 
This variation of $n$ with $P$ is smaller that that observed in a NPPM in this $P$ range, \mbox{$n_{v-\rm SiO2}^{\rm NPPM}(P) = 1.4616 + 0.0115 \times P$}~\cite{zha_acoustic_1994}. 
The value of $n_{v-\rm SiO2}^{\rm He}$ is then used to extract $v_{\rm L}$ from the more precise HRS data. 
The results are shown in Fig.~\ref{FigVelocities} together with our previous results obtained with the NPPM~\cite{ayrinhac_dynamical_2011}. 
The very small scattering of data points should be noted. 
Comparing $v_{\rm L}$ values in both media, three remarks are immediate: $(i)$ the numerical values are quite similar, $(ii)$ however, the minimum at 2 GPa with the NPPM is strongly reduced in He, and $(iii)$ the slopes above 3 GPa are very similar. As for the transverse modes, they are silent in backscattering so that the only results for $v_{\rm T}$, shown in Fig.~\ref{FigVelocities}, are from measurements in the platelet geometry. 
The new data in the NPPM are very similar to these already known from the literature~\cite{zha_acoustic_1994,polian_sound_1993}. 
As for the longitudinal mode, the numerical values for the two data sets shown are similar. 
The decrease observed between ambient pressure and 2 GPa under NPPM is reduced, but not suppressed. 

\begin{figure}
\includegraphics[width=8.5cm]{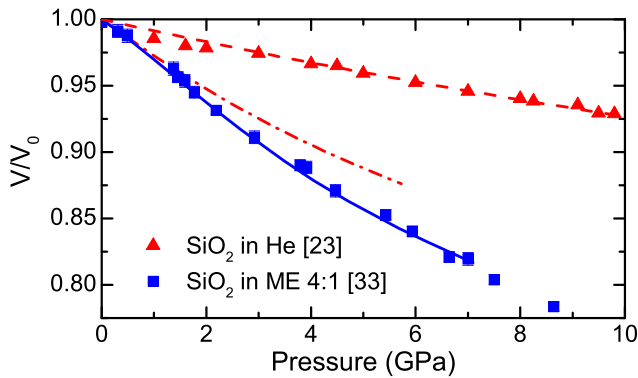}
\caption{(Color online) Relative change of volume as a function of $P$ calculated using Eq.~(\ref{EqV/V0}) for ME 4:1 (blue solid line) and He (red dashed-dotted line), compared to experimental data from literature in ME 4:1 (blue squares \cite{meade_frequency-dependent_1987}) and in He (red triangles \cite{sato_helium_2011}). 
The red dashed line is a 2$^{\rm nd}$ order Birch-Murnaghan fit~\cite{birch_finite_1947} of the volume change in He.}
\label{FigV/V0}
\end{figure}

One of the most surprising results of Refs~\cite{shen_effect_2011,sato_helium_2011} is the very small change in volume of $v$-SiO$_2$ when compressed under He. 
Such a strong effect of the pressurizing medium on the static compressibility is usually observed in materials with an open porosity, such as {\em e.g.} zeolites~\cite{hazen_zeolite_1983,chapman_guest-dependent_2008,haines_deactivation_2010}.
It results from pore filling by molecules of the pressurizing medium.
For an homogeneous {\em non-porous} material, the relative change of volume $V$ under $P$ can be calculated from the sound velocities~\cite{cook_variation_1957},
\begin{equation}
\frac{V(P)}{V_0} =\left[1+\frac{1}{\rho _0} \int_{P_0}^{P} \frac{dP}{v_{\rm L}^2\left(P\right)-\frac{4}{3} v_{\rm T}^2\left(P\right)} \right]^{-1} \:\:\:,
\label{EqV/V0}
\end{equation}
where $V_0$ and $\rho_0$ are the volume and the mass density at pressure $P_0$.
In Fig.~\ref{FigV/V0}, we plot the $V(P)$ measured for $v$-SiO$_2$ in a NPPM. 
The experimental data (squares) are in excellent agreement with the values calculated from $v_{\rm L}(P)$ and $v_{\rm T}(P)$ in ME 4:1 (solid line), as already noted by several authors (see {\em e.g.} Ref~\cite{schroeder_brillouin_1990,kondo_nonlinear_1981}). 
If Eq.~(\ref{EqV/V0}) is applied to $v$-SiO$_2$ pressurized under He using the results of Fig. \ref{FigVelocities}, the dashed-dotted curve is found. 
The disagreement with the experimental data (triangles) is obvious. 
Eq.~(\ref{EqV/V0}) fails because $\rho V \neq \rho _0 V _ 0$, which is a direct consequence of the He penetration.

\begin{figure}
\includegraphics[width=8.5cm]{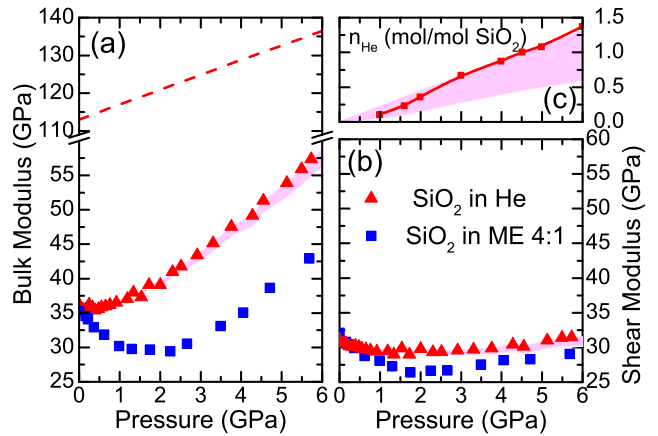}
\caption{(Color online) Elastic moduli calculated from the measured sound velocities in ME 4:1 (blue squares) and in He using the He content derived from Eq.~(\ref{Mossotti}) (red triangles): (a) bulk modulus, (b) shear modulus. 
The red dashed line in (a) is the bulk modulus derived from the Birch-Murnaghan fit shown in Fig.~\ref{FigV/V0}. 
(c) The relative He concentration $n_{\rm He}$ as a function of $P$, calculated using Eq.~(\ref{Mossotti}) (red line).
The shadowed region is delimited by the He contents calculated using the  other two hypotheses explained in the text.}
\label{FigElastMod}
\end{figure}

Turning now to the pressure dependence of the bulk modulus, the static value for $v$-SiO$_2$ in He can be derived from the volume change. This is done using a 2$^{\rm nd}$ order Birch-Murnaghan fit~\cite{birch_finite_1947} of $V(P)$ shown by the dashed line in Fig.~\ref{FigV/V0}. 
This allows calculating $B$ plotted in Fig.~\ref{FigElastMod}a. It shows a linear increase from $B=113$~GPa at $P=0$~GPa to 135~GPa at 6~GPa. 
These values are surprisingly high.
It is worth posing here to clarify the meaning of this measurement.
The static $B$ is obtained by comparing the sizes of the sample at different pressures. 
As He penetrates the network, a change in composition results from a pressure change. Hence, the static measurement of $B$ is not performed on a closed system. 
This $B$ is the effective compressibility of an open system.
It is different from the acoustic compressibility.
The BLS experiment probes the elastic moduli of the composite as a closed system. 
Indeed, the frequency of the strain wave is high, $\approx 35$~GHz, so that the He concentration in the sample cannot change over the period of the acoustic wave. 
The interest in comparing static and acoustic compressibilities is that their difference relates to the rate of He inclusion in $v$-SiO$_2$.

From the velocity measurements shown in Fig.~\ref{FigVelocities} it is possible to calculate the variation with $P$ of the elastic moduli, provided the corresponding values of the density of the medium are known. 
In the NPPM case, this is easily achieved with Eq.~\ref{EqV/V0}. 
The results are shown as squares in Fig.~\ref{FigElastMod}a and Fig.~\ref{FigElastMod}b. 
As already known, the bulk modulus $B$ decreases to a minimum around 2 GPa and increases thereafter~\cite{bridgman_effects_1953,schroeder_brillouin_1990,kondo_nonlinear_1981}. 
The shear modulus $G$ has a similar variation, with a much smaller amplitude. 
With He penetrating inside the network, the density to take into account in the equation of motion is that of a composite material made of $v$-SiO$_2$ and He atoms. 
It can be calculated if the change in volume of the sample and the amount of He are known. 
The latter can be evaluated by various methods. 
Starting from the solubility $S$ measured from ambient $P$ to 0.13 GPa~\cite{shelby_pressure_1976}, and assuming that this remains valid at higher $P$, the higher limit of the shadowed region in Fig.~\ref{FigElastMod}c is obtained. 
A second estimate is proposed in Ref.~\cite{sato_helium_2011}. 
It assumes that the volume difference between the sample under He compression and the sample compressed in NPPM is filled with He fluid in equilibrium. 
This provides the lower limit of the shadowed region in Fig.~\ref{FigElastMod}c. 
These estimates are much larger than the value 0.1~mole per mole SiO$_2$ calculated at ambient pressure from the distribution of voids in a numerical simulation~\cite{chan_theoretical_1991}.

From the measured $n_{v-\rm SiO2}^{\rm He}(P)$ with $P$, we can also derive an estimate of the amount of He entering $v$-SiO$_2$. 
To this effect we apply the Clausius-Mossotti relation to the composite sample,
\begin{equation}
3~\varepsilon _0 \frac{n^2-1}{n^2+2} = N_{\rm SiO_2}\alpha_{\rm SiO_2} + N_{\rm He}\alpha_{\rm He} \:\:\:,
\label{Mossotti}
\end{equation}
where $\varepsilon _0$ is the permittivity of free space.
Here $n=n_{v-\rm SiO2}^{\rm He}$ is the refractive index of the composite, $N_{\rm SiO_2}$ and $N_{\rm He}$ are the number densities of SiO$_2$ and He, respectively, and the $\alpha $'s are the polarizabilities. 
For He, we take $\alpha_{\rm He} = \varepsilon _0~2.602 \times 10^{-24}$~cm$^3$, approximated as constant.
$N_{\rm SiO_2}$ is available from the observed volume change in He shown in Fig.~\ref{FigV/V0}.
To estimate $\alpha_{\rm SiO_2}$, we assume that the volume calculated from $B$ measured by BLS, the dashed-dotted line in Fig.~\ref{FigV/V0}, is that occupied by the SiO$_2$ {\em skeleton} in the composite. 
Its polarizability is taken equal to that of $v$-SiO$_2$ in NPPM at the same $\rho $, known from applying the Clausius-Mossotti relation to $n_{v-\rm SiO2}^{\rm NPPM}(P)$.
Eq.~(\ref{Mossotti}) gives then $n_{\rm He} = N_{\rm He}/N_{\rm SiO_2}$, shown as a solid line in Fig.~\ref{FigElastMod}c. 
Remarkably, this line extrapolates the known solubility at low pressures~\cite{shelby_pressure_1976}, supporting this approach. 
It suggests that the He solubility remains constant over our $P$-range, leading to 1.3 mole of He per mole of SiO$_2$ at 6~GPa. 

From these various evaluations of the He content, estimates for $\rho(P)$ are obtained which are used together with the measured velocities to calculate the elastic moduli.
The triangles in Fig.~\ref{FigElastMod}a and \ref{FigElastMod}b are obtained using the $\rho(P)$ values determined from  $n_{v-\rm SiO2}^{\rm He}(P)$, as just explained.
The shadowed regions in the same figures correspond to the two limits for $\rho$ explained above. 
These shadowed regions are very narrow, as the contribution of the exact amount of He to $\rho $ is relatively minor.
$G$ is almost constant over the range of $P$. 
From measurements of the variation of the elastic constants with $P$ in crystals, it is known that compression  quite generally produces an increase of the elastic constants associated with compressive modes, while those associated with shear modes can either increase or decrease~\cite{calderon_complete_2007,jiang_elasticity_2009}. 
The conclusion is that in $v$-SiO$_2$ shear is quite insensitive to compression.

The decrease of $B$ with increasing $P$ up to $\sim$2~GPa in NPPM can be explained by interstitial voids that are progressively suppressed, and that this collapse is associated to local structural changes making the network more compliant. 
Above 2~GPa, $v$-SiO$_2$ is sufficiently compact to start behaving as a normal homogeneous solid under further compression. 
Then $B$ increases almost linearly with $P$. 
A striking result of our study is that, with He as pressurizing medium, the anomalous decrease of $B$ with increasing $P$ practically disappears. 
$B$ is nearly constant up to $\sim$1~GPa and increases thereafter. 
This indicates that, even in the first stages of compression, He penetrates in the largest voids and prevents their collapse, in line with the conclusions of earlier publications~\cite{shen_effect_2011,sato_helium_2011}. 
The structural rearrangements leading to the anomalous softening in a NPPM are no longer allowed, and the usual increase of $B$ with $P$ is then observed. 
At $P$ larger than 2~GPa, the compression of the network in presence of He is thus very similar to that observed with a NPPM.

In summary, our results demonstrate that the minima observed in $B$ and $G$ when $P$ is increased with a NPPM are suppressed when the compression is made under He.  
The $B$ minimum observed in a NPPM owing to the inter-tetrahedral flexibility is to a large extent hindered by the penetration of helium atoms in the silica network. 
It would be interesting knowing from numerical simulations how the presence of He changes the dynamics of Si$-$O$-$Si flips, or of the ``soft modes'', in the two models described in the introduction~\cite{huang_amorphous-amorphous_2004,liang_mechanical_2007,walker_origin_2007}. 
Another important result is that the acoustic bulk modulus is less than half the static one.
It must be stressed that the former is measured at constant He concentration, owing to the high frequency of the probed acoustic waves, while the later describes a composite in which the amount of He  varies.
In an open system, the static measurement underestimates the compressibility by a term due to the gas charging the composite.
The large difference found here demonstrates that He continues penetrating $v$-SiO$_2$ up to our highest investigated $P$.

\begin{acknowledgements}
The authors thank Eric Courtens for his critical and creative readings of the manuscript, as well as Julien Haines, J\'er\^ome Rouquette and Beno\^it Coasne for stimulating discussion. It is also a pleasure to thank S\'ebastien Cl\'ement and R\'emy Vialla for their help in the course of the experiments. This work was partially funded by R\'egion Languedoc-Roussillon (Omega Platform).
\end{acknowledgements}


\begin{thebibliography}{42}
\bibitem [{not()}]{note_pression}%
  \BibitemOpen
  \href@noop {} {\bibinfo  {journal} {Our measurements are performed in the
  elastic reversible domain, well below the region in which permanent
  densification occurs in a non-penetrating pressurizing medium}\ }\BibitemShut
  {NoStop}%
\end{thebibliography}
\end{document}